\documentclass[12pt, a4paper, notitlepage, oneside]{article}

\usepackage{graphicx}
\usepackage[utf8]{inputenc}
\usepackage{amsmath}
\usepackage{amssymb}
\usepackage{hyperref}
\usepackage{algorithm}
\usepackage{algpseudocode}
\usepackage{geometry}
\newgeometry{vmargin={25mm,50mm}, hmargin={35mm,35mm}}
\usepackage{fancyhdr}

\fancypagestyle{plain}{
\fancyhf{}
\pagestyle{fancy}
\fancyhead[C]{\includegraphics[height=2cm]{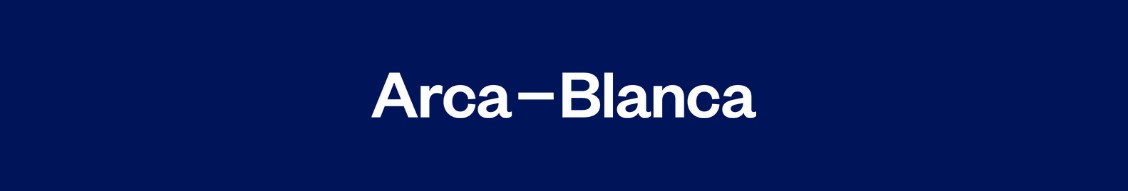}}
\fancyfoot[L]{Copyright \textcopyright \vspace{6pt} 2021 Arca Blanca}
\fancyfoot[R]{ \thepage}
}
\fancypagestyle{AB}{
\fancyhf{}
\pagestyle{fancy}
\fancyhead[L]{\includegraphics[height=1cm]{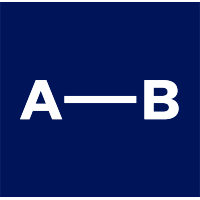}}
\fancyfoot[L]{Copyright \textcopyright \vspace{6pt} 2021 Arca Blanca}
\fancyfoot[R]{ \thepage}
}
\usepackage{authblk}
\title{Aggregating Macroeconomic Forecasts}
\author{  
  Hurwitz, Evan\\
  \texttt{evan.hurwitz@arcablanca.com}
  \and
  \v{C}evora, George\\
  \texttt{george.cevora@arcablanca.com}
  }
\affil{Arca Blanca Ltd.\\44 Maiden Lane, London WC2E 7LN}
\date{}

\title{Forecasting performance of workforce reskilling programmes}

\date{}

\begin{document}

\maketitle

\abstract{Estimating success rates for programmes aiming to reintegrate the unemployed into the workforce is essential for good stewardship of public finances. At the current moment, the methods used for this task are based on the historical performance of comparable programmes. In light of Brexit and Covid-19 simultaneously causing a shock to the labour market in the UK we developed an estimation method that is based on fundamental factors involved - workforce demand and supply - as opposed to the historical values which are quickly becoming irrelevant. With an average error of 3.9\% of the re-integration success rate, our model outperforms the best benchmark known to us by 53\%.}

\newpage
\section{Introduction}
One of the primary aims of The Department of Work and Pensions [DWP] in the United Kingdom is to help unemployed people re-enter the workforce. This aim is achieved through the \emph{reskilling} of job-seekers as appropriate given changing needs and environments and/or \emph{sourcing} jobs appropriate to the skills of the unemployed. 

The success rate (\emph{performance}) of these reskilling and sourcing programmes (\emph{programmes}) is crucial to evaluating them.  Estimating the performance of future programmes is highly valuable to both the DWP which needs to allocate limited public resources, and providers of programmes that are partially renumerated based on programme performance. The usual approach to this estimation problem is to use average historic performance of comparable projects, which 1) is blind to the varying demand and supply of workforce that are fundamental for re-integrating workers into the workforce \cite{escobari2019realism,holland2015workforce,lucas1978equilibrium}, and therefore 2) cannot react to large shocks to these factors.

The dual crises of Covid-19 and Brexit increased the need for reskilling \cite{coates2020shutdown} while at the same time provided a step-change in the demand and supply making the conventional approach unfit for purpose as the historical performance of programmes has lost its relevance for forecasting of the future.
Brexit is likely to result in lower demand for the workforce through loss of jobs \cite{erken2018measuring}, as well as decrease the supply of workforce through restrictions on migration \cite{wadsworth2016brexit}. At the same time, Brexit is forecasted to cause a significant change to the sectoral composition of the economy \cite{gasiorek2018manufacturing} creating a need for reskilling the workforce. Covid-19 too has a large, uncertain and complex potential to change both workforce supply and demand, and sectoral composition of the economy \cite{coates2020shutdown}.

 Workforce supply \cite{guinn1989changing} and demand \cite{bolton2009issues} are uncertain and ill-defined, yet still more feasible to estimate than programme performance directly. Estimating the workforce demand and supply first and subsequently integrating them into the estimation of programmes' performance, therefore, has the potential to bring significant value to both DWP and reskilling providers. 

In this paper, we describe a methodology that utilises workforce supply and demand information to improve forecasts of reskilling programmes. We demonstrate the benefits of this approach on two UK regions across 8 year period and 2 different UK geographical regions for which programme performance data was available to us.

\newpage
\section{Data and Methods}
\label{sec:data}
Data-driven models are crucially reliant on high-quality input variables. Unfortunately, the variables identified as fundamental to predicting programme performance are of low quality. For instance, unemployment figures do not accurately reflect the number of individuals looking for work as some of the unemployed are unable or unwilling to work, while others may be looking for a job without being recorded as unemployed \cite{chadi2010distinguish}. Establishing high-quality proxies for these variables is therefore crucial for the successful forecasting of programme performance.

We trained and tested our model on two undisclosed regions covered by a proprietary dataset available to us. However, we performed and provided exploratory data analysis for further 5 arbitrary regions of the UK to demonstrate statistical properties of the input variables.

\subsection{Performance}
\label{sec:performance}
For the purpose of this study, we define the performance of a programme as a proportion of people who successfully reintegrated into the workforce during the programme's duration. In order to be considered successfully reintegrated, a person must be continuously employed for at least 6 months after entering the programme, working a minimum of 16 hours per week.
We obtained this information from a proprietary dataset containing information on programme entry date for over 30000 individuals across two UK counties between the years 2011 and 2018 and whether or not they have successfully entered the workforce.

Aggregating the data by year of programme entry leaves us with 14 datapoints indicating programme performance as the last applicants to enter the programme entered in 2017. While this number is undeniably low, we believe no better dataset exists.

\subsection{Demand}
\label{sec:demand}
\begin{figure}
\centering
    \includegraphics[width=10cm]{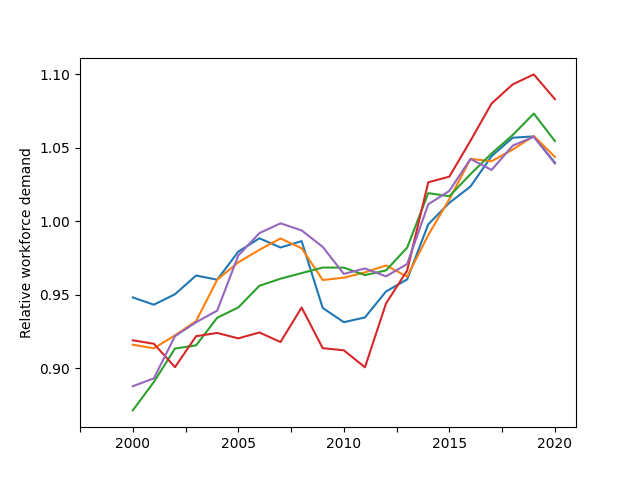}
    \caption{While the workforce demand follows broadly similar trends within 5 regions we have explored, a significant variance between regions remains. This is likely to deferentially impact performance of reskilling programs in those regions.}
    \label{fig:NormDemand}
\end{figure}

When modelling workforce demand we estimate how many people are required by businesses to meet the economic needs of the region. The best proxy for this value we are aware of are the employment figures. A dataset provided by ONS~\cite{nomisunemployment} covers years between 2000 and 2020. The employment figures are broken down by UK county and by industry sector. While the individual geographic areas follow broadly similar trends, significant variance in employment by region remains (see figure \ref{fig:NormDemand}). The split into 48 individual industry sectors is of interest as it indicates the rough level of skill required by the local economy, however, the limited size of data ruled out its use at this time.

\newpage

The indirect relationship between employment and demand for workforce required usage of change in employed population as opposed to raw employment figures as an indication of demand for workforce \cite{stephany2021online}. The change is simply approximated by a year-on-year difference.

\subsection{Supply}
\label{sec:supply}
\begin{figure}
    \centering
    \includegraphics[width=10cm]{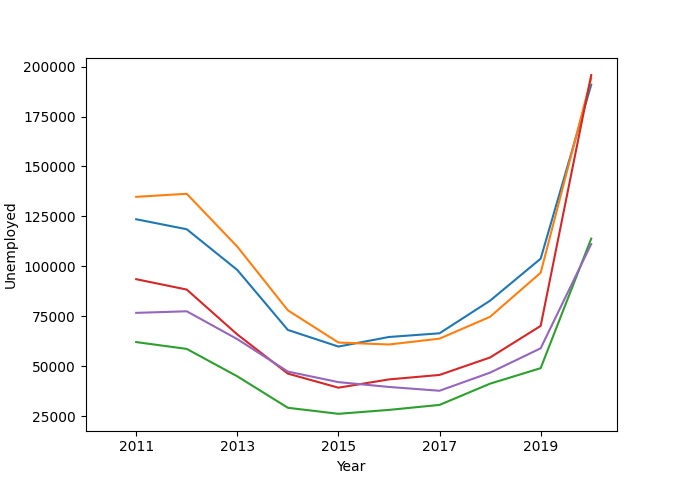}
    \caption{Similarly to demand, the unemployment also follows broadly similar trends within 5 regions we have explored, a significant variance between regions remains. This is likely to deferentially impact performance of reskilling programmes in those regions.}
    \label{fig:Unemp}
\end{figure}

\begin{figure}
    \centering
    \includegraphics[width=10cm]{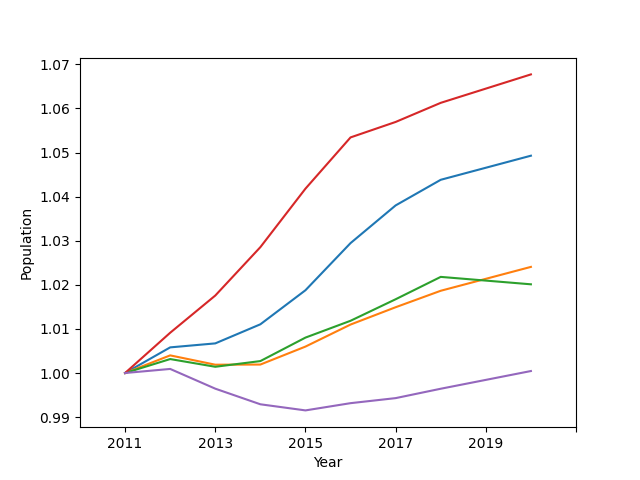}
    \caption{The population growth varies significantly between regions. This is likely to diferentially impact performance of reskilling programmes in those regions and thus is an important factor to consider when forecasting programme performance. The values are baselined by year 2011 to make the differences in growth better visible.}
    \label{fig:RawPop}
\end{figure}

When modelling workforce supply we aim to estimate how many people can take up new jobs. We base this estimation on demographic and unemployment data. 
The programmes for which we are forecasting performance are focused on benefit claimants rather than simply unemployed therefore we align our analysis with this notion of unemployment. 
The unemployment data we use was obtained from Nomis~\cite{nomisunemployment} and consists of several individuals that have been out of work for at least six months by region. 
The ONS population size dataset~\cite{onspopulation} we use is extrapolated from census data allowing for a split by age and region. The age is of particular importance as it allows us to understand the number of unemployed individuals shown in figure \ref{fig:Unemp} in the context of the size of the working population shown in figure \ref{fig:RawPop}. This leaves us with regional unemployment as a percentage of the regional working-age population, which is an attainable approximation of the available supply of labour within a region. 

\subsection{Methods}
As a result of the limited size of data, we can only use simple methods of inference, more complex methods would only improve model performance with an increase in input dimensionality, which in turn would cause the \emph{curse of dimensionality} \cite{bellman1956dynamic}.

A linear regression model is sufficient to model simple relationships such as those evaluated here. The simplicity of the model is in this case used as a constraint to avoid overfitting, which is a definite risk factor given the small number of data points. However, such a model will have a wider error margin than if further variables or more complex modelling techniques had been utilised or applied. This trade-off is acceptable as the potential loss of accuracy is preferable to overfitting which would make the model unusable.

The variables used to predict performance (see section \ref{sec:performance}) are the supply and demand measures defined in sections \ref{sec:supply} and \ref{sec:demand}. The Linear Regression model from scikit-learn was used for this purpose. This method uses unregularized least square error as its cost function.

\section{Results}
To quantify the benefit of data-driven forecasting in this area we benchmark the predictive capability of our model against the conventional approach - taking the arithmetic mean of historical performance.

\begin{figure}
    \centering
    \includegraphics[width=\linewidth]{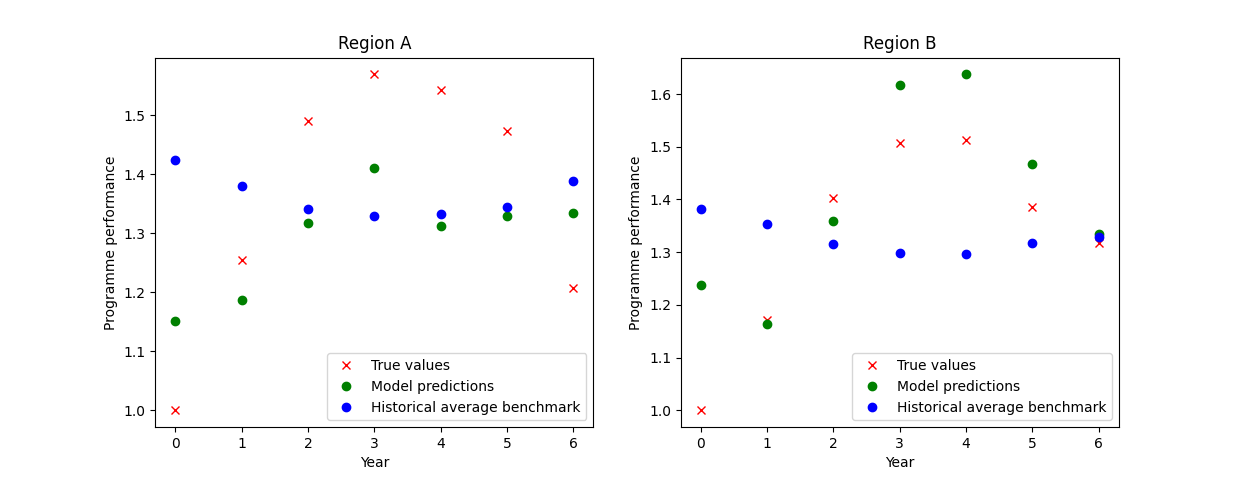}
    \caption{Our model clearly outperforms the benchmark at estimating performance of reskilling programmes with mean absolute error of 3.9 and 6 (in percent performance) respectively. The values are baselined by the first year of the programme performance for confidentiality.}
    \label{fig:eval}
\end{figure}

We have followed a conventional k-fold validation procedure where $k$ was aligned with the size of our dataset to maximise the number of samples.
The (still) low number of samples allows us to show errors on each data point in figure \ref{fig:eval}.

The mean absolute error of the model predictions is 3.9\%, compared to  6.0\% for the benchmark, meaning the benchmark is 53.8\% more inaccurate at the forecasting task. The standard deviations of the error values obtained are 4.5\% and 7\% respectively.

\newpage
\section{Discussion}
Predicting the performance of reskilling and job sourcing programmes (\emph{programmes}) is of high importance to the Department of Work and Pensions as well as the providers delivering such programmes. The current method for estimating the performance is based simply on previous performance of similar programmes, which is an estimation method that cannot react to step changes in the fundamental factors driving the performance. Given Brexit and Covid-19 crises each simultaneously causing a shock to the workforce market, there is an urgent need for an estimation method based on fundamental factors for the performance rather than just history which has been made irrelevant by these shocks.

In this paper, we introduced a method that bases performance forecasts on workforce supply and demand and demonstrated its accuracy on programme performance data available to us by benchmarking it against the current conventional method. The achieved error rate was 3.9\% of the programme's actual performance. The benchmark method produced errors more than 50\% greater demonstrating the value of data-driven forecasting as opposed to simple historical comparisons.

The dataset used consisted of 8 years of programmes' performance over an 8 year period before Brexit and Covid, with no major shocks to the labour market. We, therefore, postulate that the merit of the method we introduced in this paper should be significantly increased as historical performance is now less relevant than it was before - we expect the performance of the benchmark method to deteriorate from reported levels far faster than the model based on fundamental factors.

While the results are truly encouraging, it is necessary to acknowledge that both modelling and evaluation used extremely small datasets (14 data points). We are not aware of a larger dataset being publicly available, though a larger dataset might be obtainable through Freedom of Information Requests. Such effort is the logical next step in further improving the estimation of programmes' performance. A relatively low range of performance values have been covered by the dataset, which introduces further issue out-of-sample forecasting - a definite limitation of our estimates.

 In conclusion, data-driven modelling of the relationship between workforce demand, supply, and programme performance was shown to bring significant benefits. The predictions can be used to improve stewardship of public finances as well as inform decisions on pricing, acquisitions, expansions and tender bids with a greater degree of confidence than previously employed.

\newpage
\bibliography{bib}

\begin{thebibliography}{10}

\bibitem{bellman1956dynamic}
Richard Bellman.
\newblock Dynamic programming and lagrange multipliers.
\newblock {\em Proceedings of the National Academy of Sciences of the United
  States of America}, 42(10):767, 1956.

\bibitem{bolton2009issues}
Tom Bolton and Leonie Segal.
\newblock Issues facing the future health care workforce: the importance of
  demand modelling.
\newblock {\em Australia and New Zealand Health Policy}, 6(1), 2009.

\bibitem{chadi2010distinguish}
Adrian Chadi.
\newblock How to distinguish voluntary from involuntary unemployment: On the
  relationship between the willingness to work and unemployment-induced
  unhappiness.
\newblock {\em Kyklos}, 63(3):317--329, 2010.

\bibitem{coates2020shutdown}
Brendan Coates, Matt Cowgill, Tony Chen, and Will Mackey.
\newblock Shutdown: estimating the covid-19 employment shock.
\newblock {\em Grattan Institute Working Paper No. 2020-03}, 2020.

\bibitem{erken2018measuring}
Hugo Erken, Raphie Hayat, Carlijn Prins, Marijn Heijmerikx, and Inge de~Vreede.
\newblock Measuring the permanent costs of brexit.
\newblock {\em National Institute Economic Review}, 244:R46--R55, 2018.

\bibitem{escobari2019realism}
Marcela Escobari, Ian Seyal, and Michael Meaney.
\newblock Realism about reskilling: Upgrading the career prospects of america's
  low-wage workers. workforce of the future initiative.
\newblock {\em Center for Universal Education at The Brookings Institution},
  2019.

\bibitem{onspopulation}
Office for National~Statistics.
\newblock Population change and components of change, mid-2018 to mid-2019,
  local authorities in the uk.
\newblock online.
\newblock https://www.ons.gov.uk/visualisations/dvc1430/DD\%20map/
  component/multimap/datadownload.xlsx.

\bibitem{nomisunemployment}
Office for National~Statistics.
\newblock Claimant count by sex and age.
\newblock online, May 2021.
\newblock https://www.nomisweb.co.uk/query/construct/summary.asp?
  reset=yes\&mode=construct\&dataset=162\&version=0\&anal=1\&initsel=.

\bibitem{gasiorek2018manufacturing}
MICHAEL Gasiorek, ILONA Serwicka, and ALASDAIR Smith.
\newblock Which manufacturing sectors are most vulnerable to brexit.
\newblock {\em UK trade policy observatory. Briefing paper}, 16, 2018.

\bibitem{guinn1989changing}
Stephen~L Guinn.
\newblock The changing workforce.
\newblock {\em Training \& Development Journal}, 43(12):36--40, 1989.

\bibitem{holland2015workforce}
Brian Holland.
\newblock A workforce development systems model for unemployed job seekers.
\newblock {\em Journal of Adult and Continuing Education}, 21(2):55--76, 2015.

\bibitem{lucas1978equilibrium}
Robert~E Lucas~Jr and Edward~C Prescott.
\newblock Equilibrium search and unemployment.
\newblock In {\em Uncertainty in economics}, pages 515--540. Elsevier, 1978.

\bibitem{stephany2021online}
Fabian Stephany, Otto K{\"a}ssi, Uma Rani, and Vili Lehdonvirta.
\newblock Online labour index 2020: New ways to measure the world's remote
  freelancing market.
\newblock {\em arXiv preprint arXiv:2105.09148}, 2021.

\bibitem{wadsworth2016brexit}
Jonathan Wadsworth, Swati Dhingra, Gianmarco Ottaviano, and John Van~Reenen.
\newblock Brexit and the impact of immigration on the uk.
\newblock {\em CEP Brexit Analysis}, 5:34--53, 2016.

\end{thebibliography}
\bibliographystyle{plain}
\end{document}